\documentstyle[12pt]{article}
\def\shiftleft#1{#1\llap{#1\hskip 0.04em}}
\def\shiftdown#1{#1\llap{\lower.04ex\hbox{#1}}}

\begin{document}

\title{DIBARYONS IN NUCLEAR MATTER}
\author{Amand Faessler$^1$, A. J. Buchmann$^1$, M. I. Krivoruchenko$^{1,2}$ and\\
B. V. Martemyanov$^2$ \\
%EndAName
{\small $^1${\it Institut f\"ur Theoretische Physik, Universit\"at
T\"ubingen, Auf der Morgenstelle 14 }}\\
{\small {\it D-72076 T\"ubingen, Germany}}\\
{\small $^2${\it Institute for Theoretical and Experimental Physics,
B.Cheremushkinskaya 25}}\\
{\small {\it 117259 Moscow, Russia}}}
\date{}
\maketitle

\begin{abstract}
The possibility for occurrence of a Bose condensate of dibaryons in nuclear
matter is investigated within the framework of the Walecka model in the
mean-field approximation. Constraints for the $\omega $- and $\sigma $-meson
coupling constants with dibaryons following from the requirement of
stability of dibaryon matter against compression are derived and the effect
of $\sigma $- and $\pi $-meson exchange current contributions to the $\sigma 
$-dibaryon coupling constant is discussed. The mean-field solutions of the
model are constructed. The effective nucleon mass $m_N^{*}$ vanishes when
the density of dibaryons approaches a critical value $\rho _{DV}^{c,\max
}\approx 0.15$ fm$^{-3}$. The Green's functions of the equilibrium binary
mixture of nucleons and dibaryons are constructed by solving the
Gorkov-Dyson system of equations in the no-loop approximation. We find that
when the square of the sound velocity is positive, the dispersion laws for
all elementary excitations of the system are real functions. This indicates
stability of the ground state of the heterophase nucleon-dibaryon mixture.
In the model considered, production of dibaryons becomes energetically
favorable at higher densities as compared to estimates based on a model of
non-interacting nucleons and dibaryons.
\end{abstract}

\newpage

\section{INTRODUCTION}

\setcounter{equation}{0}

In 1977 R.Jaffe predicted \cite{Jaf} the existence of the loosely bound
dihyperon, $H$, with a mass just below the $\Lambda \Lambda $ threshold. The 
$H$ particle, if it exists decays only weakly. The exciting prospect to
observe the long-lived $H$ particle stimulated considerable theoretical and
experimental activity. Calculations of the $H$-particle mass in the bag
models \cite{ITEP, Mul}, the constituent quark model \cite{Ros, Tue}, the
Skyrme model \cite{Yos}, on the lattice \cite{Tak}, and in other models \cite
{Kun} showed that the existence of a dihyperon near the $\Lambda \Lambda $%
-threshold is plausible. The searches were proposed to examine the $H$
particle production in proton-proton \cite{Car, Bad}, proton-nucleus \cite
{San, Rot}, nucleus-nucleus \cite{Rot} collisions, via a $(K^{-},K^{+})$
reaction on a nuclear target \cite{Aer}, and through a strong decay of a $%
\Xi ^{-}$-atom system \cite{Aert}. Weak decays of the $H$ have also been
studied \cite{Krs}. The experiments \cite{Car, Aok, Rus} did not give a
positive sign for existence of the $H$ particle, however, a weak decay of
the $H$ produced in the $p-C$ reaction has been reported \cite{Shah}. The
existence of the $H$ particle remains an open question which must eventually
be settled by experiment. The candidates for double-lambda hypernucleus
whose existence constrain the binding energy of the $H$ have been observed 
\cite{Dan}.

The non-strange dibaryons with exotic quantum numbers, which have a small
width due to zero coupling to the $NN$-channel, are promising candidates for
experimental searches \cite{Mul,Kob,Glo}. The lowest-lying isospin $T=0$
dibaryons with quantum numbers $J^P=1^{-}(^1P_1)$, $3^{-}(^1F_3)$, $%
1^{+}(^3S_1,^3D_1)$, $2^{+}(^3D_2)$, etc. couple to the $NN$-channel. They
have large widths and are difficult to observe experimentally. On the other
hand, dibaryons made up of exotic quark clusters, for example, a $q^4$ and a 
$q^2$ cluster with relative orbital angular momentum $L=1$ may have unusual
quantum numbers $T=0$, $J^P=0^{-}$. The data on pion double charge exchange
(DCE) reactions on nuclei$\ $\cite{Bil}-\cite{HCl} exhibit a peculiar energy
dependence at an incident total pion energy of $190$ MeV, which can be
interpreted$\ $\cite{Mar} as evidence for the existence of a narrow $%
d^{\prime }$ dibaryon with quantum numbers $T=0$, $J^p=0^{-}$ and the total
resonance energy of $2063$ MeV. Recent experiments at TRIUMPF (Vancouver)
and CELSIUS (Uppsala) seem to support the existence of the $d^{\prime }$
dibaryon \cite{Bro}.

A method for searching narrow, exotic dibaryon resonances in the double
proton-proton bremsstrahlung reaction is discussed in Ref. \cite{Ger} and
some indications for a $d_1(1920)$ dibaryon in this reaction have recently
been found \cite{Khr}.

Dibaryons can be formed in nuclear matter. The properties of nuclear matter
with admixture of multiquark clusters are discussed by Baldin {\it et al.}$\ 
$\cite{Bal} and Chizhov {\it et al.}$\ $\cite{Chi}. In these papers, the
interaction between nucleons and multiquark clusters is included through a
van der Waals volume correction. The occurrence of a heterophase state of
nucleons and $6$-quark clusters is found to be energetically favorable in a
wide region of temperatures and densities. A model for nuclear matter with
an admixture of dibaryons with the short-range nuclear forces approximated
by a $\delta $-function-like pseudopotential is discussed by Mrowczynski 
\cite{Mro}. Nuclear matter with a Bose condensate of dibaryons belongs to a
class of heterophase substances whose properties are reviewed by Shumovskii
and Yukalov \cite{Shu}.

The occurrence of a dibaryon Bose condensate in nuclear matter results in a
softening of the equation of state (EOS) of nuclear matter. In the ideal gas
approximation, the incompressibility of the heterophase nucleon-dibaryon
matter vanishes. Occurrence of a dibaryon Bose condensate in interiors of
neutron stars decreases the maximum masses of neutron stars \cite{Kri}. The
effect of the strongly interacting $H$-particle on the structure of massive
neutron stars is investigated by Tamagaki$\ $\cite{Tam} and Olinto {\it et
al.} \cite{Hae}. The $H$-particle interaction with nucleons and the $HH$%
-intercaction are studied in the non-relativistic quark cluster model \cite
{Tue}, \cite{Str}-\cite{Sak}.

In a recent paper$\ $\cite{Buc} an exactly solvable model for a
one-dimensional Fermi-system of fermions interacting through a potential
leading to a resonance in the two-fermion channel is constructed. This model
takes the Pauli principle for fermions and the composite nature of the
two-fermion resonances into account. The behavior of the system with
increasing density can be interpreted in terms of a Bose condensation of
two-fermion resonances.

A Bose condensate of dibaryons can presumably be formed at higher densities
when relativistic effects become important. In order to describe such a
system, one should go beyond the non-relativistic many-body framework. The
relativistic field-theoretical Walecka model$\ $\cite{Wal,Chin} is known to
be very successful in describing properties of infinite nuclear matter and
of ordinary nuclei throughout the periodic table. It constitutes the basis
for the quantum hadrodynamics ({\rm QHD}) approach for studying nuclear
phenomena.

In this paper we study the effect of narrow dibaryon resonances on nuclear
matter in the framework of the Walecka model. The Lagrangian of the model
contains nucleons interacting through $\omega $- and $\sigma $ -meson
exchanges. We add to the Lagrangian dibaryons interacting with nucleons and
with each other via the exchange of $\omega $- and $\sigma $-mesons.

The outline of the paper is as follows. In the next Sect., we start with a
discussion of the properties of heterophase nucleon-dibaryon matter in the
ideal gas approximation. We give simple qualitative estimates at what
densities (chemical potentials) and for which dibaryon masses a dibaryon
Bose condensate may occur. The softening of the EOS for neutron matter and
its consequences for the gravitational stability of neutron stars is
discussed. In Sect. 3 an extension of the Walecka model including dibaryon
fields in the Lagrangian is presented. We discuss in detail the constraints
for the meson-dibaryon coupling constants which follow from the requirement
of stability of dibaryon matter against compression. We also calculate the
numerical values for these coupling constants in the additive model and
estimate the size of $\sigma $- and $\pi $-meson exchange current
corrections to the $\sigma $-dibaryon coupling constant.

In Sect. 4, the mean-field (MF) solutions to the extended Walecka model are
constructed. We show that the self-consistency equation for the effective
nucleon mass in presence of the dibaryon component can formally be reduced
to an analogous equation of the standard Walecka model. The MF solutions are
examined from the point of view of the Hugenholtz-Van Hove theorem$\ $\cite
{HV} which serves as a check for the internal consistency. It is shown that
the thermodynamic pressure coincides with the hydrostatic pressure. The
numerical results show a softening of the EOS for nuclear matter with a
dibaryon component. In Sect. 5, we construct Green's functions of the
heterophase nucleon-dibaryon matter by solving a system of the Gorkov-Dyson
equations in the no-loop approximation. We find that when square of the
sound velocity is positive, the dispersion laws for elementary excitations
are real functions, indicating the stability of the heterophase ground state
of nucleon-dibaryon matter.

\section{IDEAL GAS APPROXIMATION}

\setcounter{equation}{0}

In the ideal gas approximation, the physical picture of dibaryon
condensation is very simple. Suppose we fill a box with neutrons (see
Fig.1). Due to the Pauli principle, the neutrons occupy successively higher
energy levels. This process is continued until the chemical potential of two
neutrons on top of the Fermi sphere is larger than the dibaryon mass. When
the Fermi energy of the neutrons becomes larger, it becomes energetically
favorable for the neutrons to form dibaryons. The critical density for
dibaryon formation is determined by the mass of the lightest dibaryon. Above
the critical density, the chemical potential of the nucleons $\mu _n$ is
frozen at the value $\mu _n^{\max }=m_D/2$ where $m_D$ is the dibaryon mass.
This equation is a consequence of the chemical equilibrium with respect to
transitions $nn\leftrightarrow D.$ Dibaryons are Bose particles, so they are
accumulated in the ground state and form a Bose condensate. Dibaryons have
zero velocities, therefore they do not collide with the boundary and do not
contribute to the pressure. Because the Fermi energy of the neutrons is
frozen, the pressure does not increase with the density (see Fig. 2).
Consequently, nuclear matter loses its elasticity and its incompressibility
vanishes. Nuclear matter with such properties cannot protect neutron stars
against gravitational compression and subsequent collapse.

Obviously, a dibaryon Bose condensate does not exist in ordinary nuclei. The
equality $\mu _n=\mu _p$ follows from the charge symmetry of nuclei.
Assuming that the shell model potential for dibaryons is twice as deep as
the one for nucleons, we conclude that the masses of dibaryons coupled to
the $NN$-channel should be greater than 
\begin{equation}
\label{NC}m_D>2\mu _N=2(m_N+\varepsilon _F)=1.96\;{\rm GeV} 
\end{equation}
where $\varepsilon _F=40$ MeV is the Fermi energy of nucleons in nuclei. For
example, the $d^{\prime }$ dibaryon \cite{Bil}-\cite{HCl} is coupled to the $%
NN\pi $ channel only. In the medium, the reaction $pd^{\prime
}\leftrightarrow nnp$ is possible. In nuclei the equilibrium condition for
the chemical potential has the form $\mu _D+\mu _p=2\mu _n+\mu _p$. Since $%
\mu _p=\mu _n=\mu _N$, we arrive at the same constraint (\ref{NC}).

In chemical equilibrium with respect to the $\beta $-decays $%
p+e\leftrightarrow n+\nu _e$, the chemical potentials for neutrons and
protons in neutron matter satisfy the equality: $\mu _n=\mu _p+\mu _e$, so
that dibaryons coupled to the $nn$-, $np$-, and $pp$-channels actually occur
at slightly different densities; and dibaryon condensation starts if
dibaryon masses are smaller, respectively, than $2\mu _n$, $2\mu _n-\mu _e,$
and $2\mu _n-2\mu _e$. In case of the$\ d^{\prime }$, the critical value for
the neutron chemical potential is determined by the condition $m_{d^{\prime
}}=2\mu _n-\mu _e.$ The electron chemical potential $\mu _e$ is positive.
The $H$ particle is coupled through a double weak decay to the $NN$ channel
so that the $H$ particle occurs at $m_H=2\mu _n.$

Nonrelativistically, the density $\rho $ and pressure $p$ distributions
inside of neutron stars are described by Euler's equation 
$$
\rho \frac{\partial {\bf v}}{\partial t}+ \rho ({\bf v\cdot \nabla }){\bf v}%
=-{\bf \nabla }p-\rho {\bf \nabla }\Phi 
$$
where $\Phi $ is the gravitational potential and ${\bf v}$ is the velocity
of the neutrons. If a dibaryon Bose condensate is formed, the pressure in
the internal region of the neutron star should be constant (the dotted
horizontal line on Fig.2), and therefore ${\bf \nabla }p=0$. Gauss's law
implies $\int d{\bf S\cdot \nabla }\Phi =4\pi GM(r)$, where $M(r)$ is the
mass inside of a sphere of radius $r$. It follows that ${\bf \nabla }\Phi
\neq 0$. From Euler's equation, we get ${\bf v}$ $\neq 0$, i.e. there is no
static solution. We thus conclude that in the ideal gas approximation there
are no stable solutions if a dibaryon Bose condensate is formed and neutron
stars are gravitationally unstable. These physical arguments are also valid
in general relativity where the neutron stars are described by the
Oppenheimer-Volkoff equation \cite{Kri} and are qualitatively correct beyond
the ideal gas approximation \cite{Hae}. Therefore there is an interesting
connection between the masses of the lightest dibaryons and the upper limit
for masses of neutron stars.

The masses of several neutron stars are reliably determined to be above $%
1.4M_{\odot }$. In the tensor interaction model (a stiff model) and in the
Reid model (a soft model; for a review of these nuclear matter models see 
{\it e.g.} \cite{Sha}), the neutron chemical potential in the center of a
mass $1.4M_{\odot }$ neutron star can be evaluated to be $1090\;$MeV and $%
1015\;$MeV, respectively. The requirement that there be no Bose condensate
of dibaryons coupled to the $NN$-channel (like the $H$ -particle) inside
neutron stars gives 
\begin{equation}
\label{SC}m_D>2.18\;{\rm GeV}\;\;\;\;{\rm and\;}\;\;\;m_D>2.03\;{\rm GeV}. 
\end{equation}
These numbers are in the range of present experimental searches for
dibaryons. They are valid, however, only when the interaction of dibaryons
is neglected.

The interaction of dibaryons with neutrons and with each other increases the
pressure. The equation of state of neutron matter becomes stiffer, yielding
stability of neutron stars in some interval of densities \cite{Hae}. If
dibaryons are formed in a first order phase transition (such a scenario is
discussed by Tamagaki$\ $\cite{Tam} for the $H$-particles), neutron stars
become unstable at a critical density provided that the jump $\Delta \rho $
of the density in the phase transition point is sufficiently large \cite
{Lig,Sei}. In Fig. 2 $\Delta \rho =$ $\infty $. In Newtonian gravity, the
criterion is given by $\Delta \rho >(3/2)\rho $.

\section{DIBARYON EXTENSION OF THE WALECKA MODEL}

\setcounter{equation}{0}

Many successful phenomenological applications of QHD demonstrate that the
interactions of hadrons at large and intermediate distances can be
adequately described in terms of hadronic degrees of freedom. Many
observables are not sensitive to the contributions from very short
distances. In QHD, the effects of retardation and causality are rigorously
taken into account.

Parameters of the Walecka model are fixed by fitting the properties of
nuclear matter at the saturation density. Once the model parameters are
fixed, other consequences can be extracted without any further assumptions.
The inclusion of dibaryons to the model entails several uncertainties due to
the lack of reliable information on dibaryon masses and coupling constants.
However, many conclusions can be drawn on quite general grounds without
knowing precise values for the newly added parameters.

\subsection{Lagrangian of the model}

We consider an extension of the Walecka model by including dibaryon fields
to the Lagrangian density 
\begin{equation}
\label{III.1}
\begin{array}{c}
L=\bar \Psi (i\partial _\mu \gamma _\mu -m_N-g_\sigma \sigma
-g_\omega \omega _\mu \gamma _\mu )\Psi +\frac
12(\partial _\mu \sigma )^2-\frac 12m_\sigma ^2\sigma ^2-\frac 14F_{\mu \nu
}^2+\frac 12m_\omega ^2\omega _\mu ^2+ \\ (\partial _\mu -ih_\omega \omega
_\mu )\varphi ^{*}(\partial _\mu +ih_\omega \omega _\mu )\varphi
-(m_D+h_\sigma \sigma )^2\varphi ^{*}\varphi +{\cal L}_c. 
\end{array}
\end{equation}
Here, $\Psi $ is the nucleon field, $\omega _\mu $ and $\sigma $ are the $%
\omega $- and $\sigma $-meson fields, $F_{\mu \nu }=\partial _\nu \omega
_\mu -\partial _\mu \omega _\nu $ is the field strength tensor of the vector
field; $\varphi $ is the dibaryon field, for which we assume that it is a
isoscalar-scalar (or isoscalar-pseudoscalar) field. This assumption includes
the interesting cases of the $H$-particle and the $d^{\prime }$-dibaryon.
The values $m_\omega \ $and $m_\sigma $ are the $\omega $- and $\sigma $%
-meson masses and $g_\omega $, $g_\sigma $, $h_\omega $, $h_\sigma $ are the
coupling constants of the $\omega $- and $\sigma $-mesons with nucleons ($g$%
) and dibaryons ($h$).

The term ${\cal L}_c$ describes the conversion of dibaryons into nucleons.
The $H$-particle is coupled to the $NN$-channel through a double weak decay,
so that ${\cal L}_c=O(G_F^2).$ For the non-strange $d_1$ and the $d^{\prime
} $ dibaryon, we neglect possible virtual transitions {\it e.g.} to the $%
NN\sigma $ channel. The on-shell couplings for these dibaryons are small.
The exotic $d_1$ dibaryon decays to the $NN\gamma $-channel only, and
therefore ${\cal L}_c=O(\alpha ).$ The $d^{\prime }$ dibaryon decays to the $%
NN\pi $ channel. Due to the Adler's self-consistency condition \cite{Adl} $%
{\cal L}_c$ $\propto \partial _\mu {\bf \pi }$. In the MF approximation $%
\partial _\mu {\bf \pi }=0$, and the term ${\cal L}_c$ does not modify the
MF equations. In what follows we set ${\cal L}_c=0$. The effect of a small
term ${\cal L}_c$ reduces to providing a chemical equilibrium with respect
to transitions between dibaryons and nucleons.

The field equations corresponding to the Lagrangian have the form 
\begin{equation}
\label{III.2}(i\partial _\mu \gamma _\mu -m_N-g_\sigma \sigma
-g_\omega \omega _\mu \gamma _\mu )\Psi
=0, 
\end{equation}
\begin{equation}
\label{III.3}(-\Box -m_\sigma ^2)\sigma =g_\sigma \bar \Psi \Psi +2h_\sigma
(m_D+h_\sigma \sigma )\varphi ^{*}\varphi , 
\end{equation}
\begin{equation}
\label{III.4}((-\Box -m_\omega ^2)g_{\mu \nu }-\partial _\mu \partial _\nu
)\omega _\nu =g_\omega \bar \Psi \gamma _\mu \Psi +h_\omega \varphi ^{*}i%
\stackrel{\leftrightarrow }{\partial }_\mu \varphi -2h_\omega ^2\omega _\mu
\varphi ^{*}\varphi , 
\end{equation}
\begin{equation}
\label{III.5}((\partial _\mu +ih_\omega \omega _\mu )^2+(m_D+h_\sigma \sigma
)^2)\varphi =0. 
\end{equation}

The field operators can be expanded into $c$-number- and operator parts: 
\begin{equation}
\label{III.6}
\begin{array}{c}
\sigma =\sigma _c+\hat \sigma , \\ 
\omega _\mu =g_{\mu 0}\omega _c+\hat \omega _\mu , \\ 
\varphi =\varphi _c+\hat \varphi , \\ 
\varphi ^{*}=\varphi _c^{*}+\hat \varphi ^{*}. 
\end{array}
\end{equation}
The $c$-number parts of the fields $A=$ $\sigma $, $\omega _\mu $, $\varphi $%
, and $\varphi ^{*}$ are defined as expectation values $A_c=<A>$ over the
ground state of the system. The average values of the operator parts are
zero by definition: $<\hat A>=\;0$.

The $\sigma $-meson mean field determines the effective nucleon and dibaryon
masses in the medium 
\begin{equation}
\label{III.7}m_N^{*}=m_N+g_\sigma \sigma _c, 
\end{equation}
\begin{equation}
\label{III.8}m_D^{*}=m_D+h_\sigma \sigma _c. 
\end{equation}

The nucleon and dibaryon currents have the form 
\begin{equation}
\label{III.9}j_\mu ^N=\bar \Psi \gamma _\mu \Psi , 
\end{equation}
\begin{equation}
\label{III.10}j_\mu ^D=\varphi ^{*}i\stackrel{\leftrightarrow }{\partial }%
_\mu \varphi -2h_\omega \omega _\mu \varphi ^{*}\varphi . 
\end{equation}
The baryon number current is defined by 
\begin{equation}
\label{III.11}j_\mu ^B=j_\mu ^N+2j_\mu ^D. 
\end{equation}
The $\omega $-field is coupled to the current 
\begin{equation}
\label{III.12}j_\mu ^\omega =g_\omega j_\mu ^N+h_\omega j_\mu ^D. 
\end{equation}

The nucleon vector and scalar densities are defined by expectation values 
\begin{equation}
\label{III.13}\rho _{NV}=<\bar \Psi \gamma _0\Psi >, 
\end{equation}
\begin{equation}
\label{III.14}\rho _{NS}=<\bar \Psi \Psi >. 
\end{equation}
The scalar density of the dibaryon condensate is defined by 
\begin{equation}
\label{III.15}\rho _{DS}^c=\left| <\varphi (0)>\right| ^2. 
\end{equation}
The time evolution of the condensate $\varphi $-field is determined by the
chemical potential $\mu _D$ of dibaryons 
\begin{equation}
\label{III.16}\varphi _c(t)=e^{-i\mu _Dt}\sqrt{\rho _{DS}^c}. 
\end{equation}

It is useful to separate the contribution of the $\omega $-meson mean field
to the chemical potential energy of dibaryons 
\begin{equation}
\label{III.17}\mu _D=\mu _D^{*}+h_\omega \omega _c. 
\end{equation}
The dibaryon number density is according to Eq.(\ref{III.10}) given by 
\begin{equation}
\label{III.18}\rho _{DV}^c=2\mu _D^{*}\rho _{DS}^c. 
\end{equation}
The possibility of existence of a Bose condensate of dibaryons depends on
values of the coupling constants of dibaryons with the $\omega $- and $%
\sigma $-mesons.

\subsection{Stability of dibaryon matter against compression}

The $\omega $- and $\sigma $- meson coupling constants $h_\omega $ and $%
h_\sigma $ enter the Yukawa potential for two dibaryons 
\begin{equation}
\label{III.19}V(r)=\frac{h_\omega ^2}{4\pi }\frac{e^{-m_\omega r}}r-\frac{%
h_\sigma ^2}{4\pi }\frac{e^{-m_\sigma r}}r. 
\end{equation}
The interaction energy for dibaryons in the condensate for a constant
density distribution $\rho _D({\bf x})=$ $\rho _D$ is equal to 
\begin{equation}
\label{III.20}W=\frac 12\int d{\bf x}_1d{\bf x}_2\rho _D({\bf x}_1)\rho _D(%
{\bf x}_2)V(|{\bf x}_1-{\bf x}_2|)=2\pi N_D\rho _D(\frac{h_\omega ^2}{4\pi
m_{_\omega }^2}-\frac{h_\sigma ^2}{4\pi m_{_\sigma }^2}) 
\end{equation}
where $N_D$ is the total number of dibaryons. The integral (3.20) is linear
in density. A negative $W$ would imply instability of the system against
compression. The value $W$ is positive and the system is stable for 
\begin{equation}
\label{stab}\frac{h_\omega ^2}{4\pi m_{_\omega }^2}>\frac{h_\sigma ^2}{4\pi
m_{_\sigma }^2}. 
\end{equation}
In a nonrelativistic theory for interacting bosons \cite{Abr} and in the
model considered (see Sect.4), the requirement of stability is equivalent to
the requirement of a negative value for the boson forward scattering
amplitude and/or a positive value of square of the sound velocity ($a_s^2>0$%
). In the Born approximation, the forward scattering amplitude is, as in
Eq.(3.20), proportional to the volume integral of the potential.

The $H$-particle interactions were studied in the non-relativistic quark
cluster model \cite{Tue}, \cite{Str}-\cite{Sak} which is successful in
describing the $NN$-phase shifts. The calculation of the interaction
integral (\ref{III.20}) with the adiabatic $HH$-potential \cite{Sak} gives a
negative number, so that the $H$-dibaryon matter is probably unstable
against compression. The coupling constants of the mesons with the $H$%
-particle can be fixed by fitting the depth and the position of the minimum
of the $HH$-potential to give $h_\omega ^2=603.7$ and $h_\sigma ^2=279.2.$
These values, while yielding a negative $W$, probably overestimate the
repulsion between $H$-particles at small distances, and therefore
underestimate $\left| W\right| $. The meson-dibaryon coupling constants for
the case of the $d_1$ and $d^{^{\prime }}$ dibaryons are presently unknown.

\subsection{Coupling constants in the additive model}

In the additive picture, mesons interact with the constituents of the
dibaryon (Fig.3). For non-strange dibaryons coupled to the $NN$-channel, the 
$\sigma $ - and $\omega $- meson couplings are in the nonrelativistic
approximation simply twice the corresponding meson-nucleon coupling
constants: $h_\omega =2g_\omega $ and $h_\sigma =2g_\sigma $. The scalar
charge is, however, suppressed by the Lorentz factor. This effect decreases
the value of $h_\sigma $. Note that the magnitudes of the meson coupling
constants for the $H$-particle, extracted from the adiabatic $HH$-potential 
\cite{Sak} are consistent with the additive estimates: $h_\omega /(2g_\omega
)=0.89$ and $h_\omega /(2g_\sigma )=0.80$.

For the standard set$\ $of parameters of the Walecka model \cite{Wal,Ser92}, 
$m_\sigma =520$ MeV, $g_\omega ^2=190.4$, and $g_\sigma ^2=109.6$, the
inequality (\ref{stab}) 
\begin{equation}
\label{III.22}98.85(\frac{h_\omega }{2g_\omega })^2{\rm GeV}^{-2}>129.0(%
\frac{h_\sigma }{2g_\sigma })^2{\rm GeV}^{-2}
\end{equation}
is not satisfied. The precision of the additive estimates is, however, not
better than $30\%$ of the central values. Exchange current contributions
which violate the additivity are discussed below.

The violation of the inequality (\ref{stab}) for the additive estimates of
the meson-dibaryon couplings is not accidental. For homophase nuclear
matter, $g_\omega \ $should be greater than $g_\sigma $ in order to get
sufficient repulsion between nucleons at small distances. To reproduce the
properties of nuclear matter at the saturation density, the following
inequality should hold 
\begin{equation}
\label{III.23}\frac{g_\omega ^2}{4\pi m_{_\omega }^2}<\frac{g_\sigma ^2}{%
4\pi m_{_\sigma }^2}.
\end{equation}
At small densities, the inequality (\ref{III.23}) provides a negative value
for the interaction integral (3.20) between the nucleons, resulting to the
local instability of nuclear matter against compression. When the density
increases, the relativistic effects become important. The scalar density of
nucleons increases slower than the vector density, since it is suppressed by
the Lorentz factor $<1/\gamma >$. This finally leads to an equilibrium at
the saturation density of nuclear matter.

For exotic dibaryons like the $d^{\prime }$, one should take into account
the presence of a pion in the resonance wave function. The $\omega $-meson
decays into three pions. It is not coupled to the pion in the $d^{\prime }$
wave function (see Fig.3), so $h_\omega =2g_\omega .$

The $\sigma -{\bf \pi }$ cubic couplings are described by the effective
Lagrangian density 
\begin{equation}
\label{III.24}\Delta {\cal L}=-\frac \kappa 6\sigma ^3-\frac{\kappa ^{\prime
}}6\sigma {\bf \pi }^2 
\end{equation}
where ${\bf \pi }$ is the pion field. The cubic terms generate three-body
forces between nucleons. Phenomenological fits to the bulk nuclear
properties give$\ $\cite{Ser92} 
\begin{equation}
\label{III.25}\frac \kappa {m_N}=0.9\div 5.3. 
\end{equation}
The lower and upper values correspond, respectively, to a small negative and
a large positive term $\lambda \sigma ^4$ in the effective Lagrangian.

The non-linearities of the linear sigma-model are qualitatively different
[25] 
\begin{equation}
\label{III.26}\frac \kappa {m_N}=\frac{\kappa ^{\prime }}{m_N}=-3g_\sigma 
\frac{m_\sigma ^2-m_\pi ^2}{m_N^2}=-8.9.
\end{equation}
In the additive model for $d^{\prime }$, we get 
\begin{equation}
\label{III.27}\frac{h_\sigma }{2g_\sigma }=1+\frac{\kappa ^{\prime }}{%
24g_\sigma m_\pi }=1+0.027\frac{\kappa ^{\prime }}{m_N}.
\end{equation}
The last term describes the contribution from the first of the diagrams in
Fig.3(b). For $\left| \kappa ^{\prime }/m_N\right| <10$ the correction to
the additive value is smaller than $30\%$. The sign, however, is not defined.

The Brown-Rho scaling$\ $\cite{Bro} for non-strange hadron masses is
reproduced at the tree level for $h_\sigma =(m_D/m_N)g_\sigma $. In such a
case, 
\begin{equation}
\label{III.28}\frac{m_N^{*}}{m_N}=\frac{m_D^{*}}{m_D}. 
\end{equation}
Since $m_D>2m_N,$ Brown-Rho scaling gives $h_\sigma =(m_D/m_N)g_\sigma
>2g_\sigma .$

\subsection{Exchange current contributions to the coupling constants}

The exchange current contributions to the $\sigma $-dibaryon coupling
constants shown in Fig.4 can be extracted from the Lagrangian (3.24). The $%
\sigma $-meson field generated by a pointlike source of charge $h_\sigma $
is determined from the equation 
\begin{equation}
\label{III.29}(\Delta -m_\sigma ^2)\sigma ({\bf x})=h_\sigma \delta ({\bf x}%
). 
\end{equation}
The sign of the right hand side of the equation is chosen such as to yield a
negative (attractive) $\sigma $-meson field for a positive $h_\sigma $. It
is useful to compare Eq.(3.29) with the static limit of the equation of
motion for the $\sigma $-meson field determined by the Lagrangian ${\cal L}%
+\Delta {\cal L}$ with no dibaryon component: 
\begin{equation}
\label{III.30}(\Delta -m_\sigma ^2)\sigma ({\bf x})=g_\sigma
\sum_{k=1}^2\bar \Psi _k({\bf x})\Psi _k({\bf x})+\frac \kappa 2\sigma ^2(%
{\bf x})+\frac{\kappa ^{\prime }}6{\bf \pi }^2({\bf x}). 
\end{equation}
The constant $h_\sigma $ measures the scalar charge generating the $\sigma $%
-meson field around dibaryons, so that we can write 
\begin{equation}
\label{III.31}h_\sigma =\int d{\bf x}[g_\sigma \sum_{k=1}^2\bar \Psi _k({\bf %
x})\Psi _k({\bf x})+\frac \kappa 2\sigma ^2({\bf x})+\frac{\kappa ^{\prime }}%
6{\bf \pi }^2({\bf x})]. 
\end{equation}

Nonrelativistically, the $\sigma $-meson field created by nucleons located
at points ${\bf x}_k$ is 
\begin{equation}
\label{III.32}\sigma ({\bf x})=-\sum_{k=1}^2g_\sigma \frac{e^{-m_\sigma |%
{\bf x}-{\bf x}_k|}}{4\pi |{\bf x}-{\bf x}_k|}. 
\end{equation}
Substituting this expression into Eq.(3.31) and omitting diagonal terms, we
get for the exchange-current contribution an expression 
\begin{equation}
\label{III.33}\Delta h_\sigma (D)^{\sigma -MEC}=\frac \kappa {16\pi m_\sigma
}g_\sigma ^2<e^{-m_\sigma |{\bf x}_1-{\bf x}_2|}> 
\end{equation}
where ${\bf x}_1$ and ${\bf x}_2$ are coordinates of the two nucleons. The
exponential term should be averaged over the dibaryon wave function.

To give an order-of-magnitude estimate, we use for the relative wave
function of two nucleons Hulthen wave functions 
\begin{equation}
\label{III.34}\psi (r)=\sqrt{\frac{\mu \nu (\mu +\nu )}{2\pi (\nu -\mu )^2}}%
\frac 1r(e^{-\mu r}-e^{-\nu r}).
\end{equation}
with $\mu =1$\ fm$^{-1}$ and $\nu =4\mu .$ The correction (3.33) can be
estimated to be 
$$
\frac{\Delta h_\sigma (D)^{\sigma -MEC}}{2g_\sigma }=0.040\frac \kappa
{m_N}. 
$$

The $d^{\prime }$-dibaryon contains a constituent $\pi $-meson. The $\sigma $%
-meson field created by the pion is given by 
\begin{equation}
\label{III.35}\sigma ({\bf x})=-\frac{\kappa ^{\prime }}{12m_\pi }\frac{%
e^{-m_\sigma |{\bf x}-{\bf x}_3|}}{4\pi |{\bf x}-{\bf x}_3|} 
\end{equation}
where ${\bf x}_3$ is the pion coordinate. The total change in the $\sigma $
-meson coupling with the $d^{\prime }$ equals 
\begin{equation}
\label{III.36}\Delta h_\sigma (d^{\prime })^{\sigma -MEC}=\frac \kappa
{16\pi m_\sigma }g_\sigma \left[ g_\sigma <e^{-m_\sigma |{\bf x}_1-{\bf x}%
_2|}>+\frac{\kappa ^{\prime }}{6m_\pi }<e^{-m_\sigma |{\bf x}_1-{\bf x}%
_3|}>\right] . 
\end{equation}
Assuming that the two-body probability densities are given by the square of
the wave function (3.34), we get 
$$
\frac{\Delta h_\sigma (d^{\prime })^{\sigma -MEC}}{2g_\sigma }=0.040\frac
\kappa {m_N}(1+0.108\frac{\kappa ^{\prime }}{m_N}). 
$$

The contribution of the pion exchange currents can be evaluated in a similar
way. The pion field created by nucleons has the form 
\begin{equation}
\label{III.37}{\bf \pi }({\bf x})=\sum_{k=1}^2i\frac{g_\sigma }{2m_N}{%
\hbox{$\tau$}\llap{\hbox{$\tau$}\hskip 0.04em}%
\llap{\lower.04ex\hbox{\shiftleft{\hbox{$\tau$}}}}}_k({\hbox{$\sigma$}%
\llap{\hbox{$\sigma$}\hskip 0.04em}%
\llap{\lower.04ex\hbox{\shiftleft{\hbox{$\sigma$}}}}}_k\cdot {\bf \nabla )}%
\frac{e^{-m_\pi |{\bf x}-{\bf x}_k|}}{4\pi \left| {\bf x}-{\bf x}_k\right| }%
. 
\end{equation}
For dibaryons consisting of two nucleons and also for the $d^{\prime }$, the
exchange pion current contributions are given by 
\begin{equation}
\label{III.38}\Delta h_\sigma (D,d^{\prime })^{\pi -MEC}=\frac{\kappa
^{\prime }}{144\pi }(\frac{g_\sigma }{2m_N})^2({\hbox{$\tau$}%
\llap{\hbox{$\tau$}\hskip 0.04em}%
\llap{\lower.04ex\hbox{\shiftleft{\hbox{$\tau$}}}}}_1\cdot {\hbox{$\tau$}%
\llap{\hbox{$\tau$}\hskip 0.04em}%
\llap{\lower.04ex\hbox{\shiftleft{\hbox{$\tau$}}}}}_2)({\hbox{$\sigma$}%
\llap{\hbox{$\sigma$}\hskip 0.04em}%
\llap{\lower.04ex\hbox{\shiftleft{\hbox{$\sigma$}}}}}_1\cdot {\hbox{$\sigma$}%
\llap{\hbox{$\sigma$}\hskip 0.04em}%
\llap{\lower.04ex\hbox{\shiftleft{\hbox{$\sigma$}}}}}_2)\left\langle \left(
\frac 2{\left| {\bf x}_1-{\bf x}_2\right| }-m_\pi \right) e^{-m_\pi |{\bf x}%
_1-{\bf x}_2|}\right\rangle . 
\end{equation}
The evaluation of this expression using the wave function (3.34) gives 
$$
\frac{\Delta h_\sigma (d^{\prime })^{\pi -MEC}}{2g_\sigma }=0.0015\frac{%
\kappa ^{\prime }}{m_N}({\hbox{$\tau$}\llap{\hbox{$\tau$}\hskip 0.04em}%
\llap{\lower.04ex\hbox{\shiftleft{\hbox{$\tau$}}}}}_1\cdot {\hbox{$\tau$}%
\llap{\hbox{$\tau$}\hskip 0.04em}%
\llap{\lower.04ex\hbox{\shiftleft{\hbox{$\tau$}}}}}_2)({\hbox{$\sigma$}%
\llap{\hbox{$\sigma$}\hskip 0.04em}%
\llap{\lower.04ex\hbox{\shiftleft{\hbox{$\sigma$}}}}}_1\cdot {\hbox{$\sigma$}%
\llap{\hbox{$\sigma$}\hskip 0.04em}%
\llap{\lower.04ex\hbox{\shiftleft{\hbox{$\sigma$}}}}}_2). 
$$
The pion exchange current correction to the $\sigma $-dibaryon coupling is
therefore significantly smaller than the corresponding sigma exchange
current corrections.

In view of the estimates (3.25) and (3.26) the exchange current corrections
cannot exceed $30\%-40\%$ of the additive values. At present, the coupling
constants of the mesons with dibaryons are not known with sufficient
precision to draw any definite conclusion concerning the stability of
dibaryon matter.

The $H$-particle, however, is studied better than other dibaryons. In what
follows, we use a realistic $HH$-interaction, based on the quark cluster
model \cite{Tue}, \cite{Str}-\cite{Sak}.

\section{THE MEAN-FIELD SOLUTIONS}

\setcounter{equation}{0}

The MF solutions of the Walecka model are asymptotically exact in the high
density limit \cite{Ser92}. They serve as a starting point for the
calculation of corrections for finite densities within the QHD approach.

Neglecting the operator parts of the meson fields in Eqs.(3.2)-(3.5), we get
the following expressions for the meson mean fields 
\begin{equation}
\label{IV.1}\omega _c=\frac{g_\omega \rho _{NV}+h_\omega 2\mu _D\rho _{DS}^c%
}{\tilde m_\omega ^2}=\frac{g_\omega \rho _{NV}+h_\omega 2\mu _D^{*}\rho
_{DS}^c}{m_\omega ^2}, 
\end{equation}
\begin{equation}
\label{IV.2}\sigma _c=-\frac{g_\sigma \rho _{NS}+h_\sigma 2m_D\rho _{DS}^c}{%
\tilde m_\sigma ^2}=-\frac{g_\sigma \rho _{NS}+h_\sigma 2m_D^{*}\rho _{DS}^c%
}{m_\sigma ^2}. 
\end{equation}
The effective masses of mesons in a heterophase nucleon-dibaryon matter are
given by 
\begin{equation}
\label{IV.3}\tilde m_\sigma ^2=m_\sigma ^2+2h_\sigma ^2\rho _{DS}^c, 
\end{equation}
\begin{equation}
\label{IV.4}\tilde m_\omega ^2=m_\omega ^2+2h_\omega ^2\rho _{DS}^c. 
\end{equation}
The mechanism responsible for the change of meson masses is essentially the
Higgs mechanism.

Substituting expression (3.16) into Eq.(3.5), we get 
\begin{equation}
\label{IV.5}\mu _D^{*}=m_D^{*}. 
\end{equation}

The nucleon and dibaryon chemical potentials have the form 
\begin{equation}
\label{IV.6}\mu _N=E_F^{*}+g_\omega \omega _c, 
\end{equation}
\begin{equation}
\label{IV.7}\mu _D=m_D^{*}+h_\omega \omega _c, 
\end{equation}
where $E_F^{*}=\sqrt{m_N^{*2}+{\bf k}_F^2}$ is the Fermi energy of nucleons
with the effective mass $m_N^{*}$.

\subsection{The self-consistency equation}

The nucleon vector and scalar densities are given by 
\begin{equation}
\label{IV.8}\rho _{NV}=\gamma \int \frac{d{\bf k}}{(2\pi )^3}\theta
(k_F-\left| {\bf k}\right| ), 
\end{equation}
\begin{equation}
\label{IV.9}\rho _{NS}=\gamma \int \frac{d{\bf k}}{(2\pi )^3}\frac{m_N^{*}}{%
E_{}^{*}({\bf k})}\theta (k_F-\left| {\bf k}\right| ). 
\end{equation}
The statistical factor $\gamma =4$ $(2)$ for nuclear (neutron) matter.

The self-consistency equation has the form 
\begin{equation}
\label{IV.10}m_N^{*}=m_N-\frac{g_\sigma }{m_\sigma ^2}(g_\sigma \rho
_{NS}+h_\sigma \rho _{DV}^c). 
\end{equation}
It follows from Eqs.(3.7), (3.18), (4.2), and (4.5). The total baryon number
density equals $\rho _{TV}=\rho _{NV}+2\rho _{DV}^c.$ Eq.(4.10) can be
transformed to a form equivalent to the self-consistency equation of the
standard Walecka model without dibaryons: 
\begin{equation}
\label{IV.11}m_N^{*}=\tilde m_N-\frac{g_\sigma ^2}{m_\sigma ^2}\rho _{NS}, 
\end{equation}
where 
\begin{equation}
\label{IV.12}\tilde m_N=m_N\frac{\rho _{DV}^{c,\max }-\rho _{DV}^c}{\rho
_{DV}^{c,\max }}, 
\end{equation}
\begin{equation}
\label{IV.13}\rho _{DV}^{c,\max }=\frac{m_Nm_\sigma ^2}{g_\sigma h_\sigma }%
=0.1507(\frac{2g_\sigma }{h_\sigma }){\rm fm}^{-3}. 
\end{equation}
If the densities $\rho _{TV}$ and $\rho _{DV}^c$ are fixed, equation (4.11)
allows to find the effective nucleon mass $m_N^{*}$. Solutions to Eq.(4.11)
exist for an arbitrary total density $\rho _{TV}$ when the value $\tilde m_N$
is positive. This is the case for $\rho _{DV}^c<\rho _{DV}^{c,\max }$. Note
that the dibaryon mass does not enter Eq.(4.11) directly.

In Figs.5 (a,b) we show for $h_\omega =2g_\omega $ the critical density for
occurrence of a Bose condensate of dibaryons in (a) symmetric nuclear matter
and (b) neutron matter as a function of the $\sigma $-dibaryon coupling
constant. The critical density is determined from equation $2\mu _N-\mu
_D=0. $ In the region $a_s^2>0$ the requirement (3.21) is fulfilled. The
dibaryon components of the heterophase nuclear- and neutron-dibaryon matter
are stable against compression.

For $2m_N\leq m_D\leq 1.89\;$GeV, we start at zero density from heterophase
nuclear-dibaryon matter $(a)$. With increasing density, the matter
transforms into homophase nuclear matter and then again back to heterophase
nuclear-dibaryon matter. For $m_D>1.89\;$GeV, we start at zero density from
homophase nuclear matter which converts with increasing density (at $\rho
_{TV}>\rho _0$ for $h_\sigma /(2g_\sigma )<0.8754$) into heterophase
nuclear-dibaryon matter.

When the $\sigma $-meson coupling with dibaryons decreases, the effective
dibaryon mass $m_D^{*}$ increases (see Eq.(3.8)) and dibaryon formation is
therefore suppressed. We thus conclude that the curves in Figs.5 (a,b)
should have a negative slope for a transition from homophase to heterophase
matter and a positive slope for a transition from heterophase to homophase
matter. The region of instability for neutron matter is greater because at
the same density we have higher values of the neutron chemical potentials
and therefore more favorable conditions for the production of dibaryons.

We show the results for $d_1$-dibaryon \cite{Khr} with possible quantum
numbers $J^p=1^{+}$. For such dibaryons, the MF equations derived above for
the scalar (pseudoscalar) dibaryons $J^p=0^{\pm }$ remain the same. In case
of the $J^p=1^{\pm }$ dibaryons, the dibaryon part of the Lagrangian (\ref
{III.1}) can be written in the form 
$$
{\cal L}_D=-D_\mu ^{*}\varphi _\nu ^{*}D_\mu \varphi _\nu +(m_D+h_\sigma
\sigma )^2\varphi _\mu ^{*}\varphi _\mu +\lambda D_\mu ^{*}\varphi _\nu
^{*}+\lambda ^{*}D_\mu \varphi _\nu 
$$
where $D_\mu =\partial _\mu +ih_\omega \omega _\mu $ and $\lambda $ and $%
\lambda ^{*}$ are Lagrange multipliers. The field equations are 
$$
\begin{array}{c}
(D_\nu ^2+(m_D+h_\sigma \sigma )^2)\varphi _\mu -D_\mu \lambda =0, \\ 
D_\mu \varphi _\mu =0. 
\end{array}
$$
The last equation removes the unphysical timelike component of the dibaryon
vector field in the co-moving frame of the particle. It is equivalent to the
requirement $u_\mu \varphi _\mu =0$ where $u_\mu $ is a four-vector velocity
of the particle.

The condensate field evolves like%
$$
<\varphi _\mu (t)>=(0,{\bf e})\sqrt{\rho _{DS}^c}e^{-i\mu _Dt} 
$$
where ${\bf e}$ is a unit vector. The field equations give $\lambda =\lambda
^{*}=0$ and result in Eqs.(4.1) and (4.2). The energy and pressure have the
form (4.22) and (4.23) of the $J^p=0^{\pm }$ dibaryons.

In Fig.6 we classify possible behaviors of the difference $2\mu _N-\mu _D$
between the chemical potentials of nucleons and dibaryons with growth of the
dibaryon fraction.

It is clear that when the difference is positive and the dibaryon density is
zero, $\rho _{DV}^c=0$, production of dibaryons is energetically favorable.
In such a case, the fraction of dibaryons increases. The state $\rho
_{DV}^c=0$ is therefore unstable. If the difference $2\mu _N-\mu _D$ is
negative and the substance consists of dibaryons only, production of
nucleons is energetically favorable. This state is unstable. If the
difference $2\mu _N-\mu _D$ is zero, but increases with the dibaryon
fraction, small fluctuations bring the substance away from the equilibrium.
Such a state is unstable also.

The system is stable in the following three cases.

(i) Homophase nuclear and neutron matter: 
\begin{equation}
\label{IV.14}
\begin{array}{c}
2\mu _N-\mu _D<0, \\ 
\rho _{DV}^c=0. 
\end{array}
\end{equation}

(ii) Homophase dibaryon matter: 
\begin{equation}
\label{IV.15}
\begin{array}{c}
2\mu _N-\mu _D>0, \\ 
2\rho _{DV}^c=\rho _{TV}. 
\end{array}
\end{equation}

(iii)\ Heterophase nucleon-dibaryon matter: 
\begin{equation}
\label{cheq}2\mu _N-\mu _D=0, 
\end{equation}
\begin{equation}
\label{poen}\frac{d(2\mu _N-\mu _D)}{d\rho _{DV}^c}\left| _{\rho
_{TV}}\right. <0. 
\end{equation}
In the first case there are no dibaryons, in the second case there are no
nucleons, and in the third case we have a heterophase mixture of nucleons
and dibaryons. Small fluctuations around the state $2\mu _N-\mu _D=0$ bring
the substance back to the equilibrium point. Eqs.(4.16) and (4.17) therefore
describe a stable equilibrium.

In Fig.7 (a) we show the nucleon effective mass $m_N^{*}$ versus the
dibaryon fraction $2\rho _{DV}^c/\rho _{TV}$ in heterophase nuclear matter
and in heterophase neutron-dibaryon matter for the coupling constants $%
h_\omega =2g_\omega $ and $h_\sigma /(2g_\sigma )=0.8$. At the same total
baryon number density, neutron matter is more relativistic. The scalar
charge density of neutron matter is thus lower, the $\sigma $-meson mean
field is smaller, and therefore the effective nucleon mass is greater. On
the plot, the dotted lines corresponding to neutron matter lie above the
solid lines corresponding to symmetric nuclear matter.

When the dibaryon vector density $\rho _{DV}^c$ approaches its maximum value 
$\rho _{DV}^{c,\max },$ the effective nucleon mass vanishes. This effect can
be interpreted as follows. Two nucleons on the top of the Fermi sphere have
energy $2E_F^{*}$. In chemical equilibrium with respect to transitions $%
NN\leftrightarrow D$, the relation $2E_F^{*}=m_D^{*}$ holds. In transitions $%
NN\leftrightarrow D$ the baryon vector charge does not change. However, the
scalar charge changes. For two nucleons the scalar charge equals $2g_\sigma 
\frac{m_N^{*}}{E_F^{*}}$, whereas a dibaryon in the condensate has scalar
charge $h_\sigma $. When the system is nonrelativistic, formation of new
dibaryons is accompanied by a decrease of the scalar charge density, since $%
2g_\sigma \frac{m_N^{*}}{E_F^{*}}\approx 2g_\sigma >1.6g_\sigma =h_\sigma $.
This phenomenon is reflected in the slight increase of the effective nucleon
mass with the dibaryon fraction at low total baryon number densities. When
the density is high, the system becomes relativistic, and so the Lorentz
factor $\frac{m_N^{*}}{E_F^{*}}$ comes into play. The scalar charge of the
two nucleons is small, whereas the dibaryon scalar charge is large. As a
result, the scalar charge density increases, the scalar mean field
increases, and the effective nucleon mass decreases with the dibaryon
fraction. In the standard Walecka model, the effective nucleon mass vanishes
at infinite density. In the model considered, the effective nucleon mass
vanishes when the density of dibaryons approaches the value $\rho
_{DV}^{c,\max }$. Note that the behavior of the effective nucleon mass $%
m_N^{*}$ with growing dibaryon fraction, shown in Fig.7 (a), does not depend
on the vacuum value of the dibaryon mass, because the dibaryon mass does not
enter the self-consistency equation (4.11).

In Fig.7 (b) we show the difference for the chemical potentials versus the
dibaryon fraction for $m_D=1.96$ GeV. Since the vacuum dibaryon mass does
not enter the self-consistency equation (4.11) and enters linearly in the
difference $2\mu _N-\mu _D$ (see Eqs.(3.8) and (4.6),(4.7)), the results for
other dibaryon masses can be obtained simply by vertical parallel
displacements of the curves. The results for the $m_D=2.08$ GeV ($d^{\prime
} $ dibaryon) can be obtained {\it e.g. }by a $100$ ${\rm MeV}$ negative
shift, etc.

The MF solutions exist at all densities $\rho _{TV}$ for sufficiently small
densities of dibaryons, $\rho _{DV}^c$ $<\rho _{DV}^{c,\max }$. It means
that we can always investigate the stability of homophase nuclear and
neutron matter with respect to formation of a dibaryon Bose condensate. When
the total density $\rho _{TV}$ is very high, the dibaryon production is
energetically favorable. The MF solutions disappear, however, before the
system reaches equilibrium.

\subsection{Thermodynamic consistency checks for the mean-field solutions}

The canonical energy-momentum tensor corresponding to the Lagrangian density
(2.1) can be written in the form 
$$
T_{\mu \nu }=T_{\mu \nu }^N+T_{\mu \nu }^\sigma +T_{\mu \nu }^\omega +T_{\mu
\nu }^D 
$$
where 
\begin{equation}
\label{IV.18}T_{\mu \nu }^N=\bar \Psi i\gamma _\mu \partial _\nu \Psi , 
\end{equation}
\begin{equation}
\label{IV.19}T_{\mu \nu }^\sigma =-\frac 12g_{\mu \nu }(\partial _\tau
\sigma \partial _\tau \sigma -m_\sigma ^2\sigma ^2)+\partial _\mu \sigma
\partial _\nu \sigma , 
\end{equation}
\begin{equation}
\label{IV.20}T_{\mu \nu }^\omega =\frac 12g_{\mu \nu }((\partial _\tau
\omega _\lambda -\partial _\lambda \omega _\tau )\partial _\tau \omega
_\lambda -m_\omega ^2\omega _\tau \omega _\tau )-\partial _\mu \omega _\tau
\partial _\nu \omega _\tau 
\end{equation}
\begin{equation}
\label{IV.21}T_{\mu \nu }^D=2\partial _\mu \varphi ^{*}\partial _\nu \varphi
-h_\omega \omega _\mu \varphi ^{*}i\stackrel{\leftrightarrow }{\partial }%
_\nu \varphi . 
\end{equation}
The energy density $\varepsilon =<T_{00}>$ given by average value of the $%
T_{00}$ component has the form 
\begin{equation}
\label{IV.22}\varepsilon =\gamma \int_0^{k_F}\frac{d{\bf k}}{(2\pi )^3}%
(E^{*}({\bf k})+g_\omega \omega _c)+(m_D^{*}+h_\omega \omega _c)\rho
_{DV}^c+\frac 12m_\sigma ^2\sigma _c^2-\frac 12m_\omega ^2\omega _c^2. 
\end{equation}
Here, $E^{*}({\bf k})+g_\omega \omega _c$ is the total energy of a nucleon
with momentum ${\bf k}$, and $m_D^{*}+h_\omega \omega _c$ is the total
energy of a dibaryon in the ground state (with zero total momentum) in the
external $\omega $-meson mean field. The last two terms are the
contributions of the classical $\omega $- and $\sigma $-meson fields to the
energy density.

The hydrostatic pressure $p=-\frac 13<T_{ii}>$ has the form 
\begin{equation}
\label{IV.23}p=\frac \gamma 3\int_0^{k_F}\frac{d{\bf k}}{(2\pi )^3}\frac{%
{\bf k}^2}{E^{*}({\bf k})}-\frac 12m_\sigma ^2\sigma _c^2+\frac 12m_\omega
^2\omega _c^2. 
\end{equation}
Because dibaryons in the condensate are at rest they do not contribute to
the pressure.

In agreement with the general requirements 
\begin{equation}
\label{IV.24}\mu _N=\frac{\partial \varepsilon }{\partial \rho _{NV}}, 
\end{equation}
\begin{equation}
\label{IV.25}\mu _D=\frac{\partial \varepsilon }{\partial \rho _{DV}^c}. 
\end{equation}
The pressure can be calculated in two different ways: from Eq.(4.23) and
from the thermodynamic relation 
\begin{equation}
\label{IV.26}p=-\varepsilon +\mu _N\rho _{NV}+\mu _D\rho _{DV}^c. 
\end{equation}
It is not difficult to check that the thermodynamic pressure (4.26)
coincides with the hydrostatic pressure (4.23).

The Hugenholtz-Van Hove theorem$\ $\cite{HV} (HV) requires that the energy
of fermions at the Fermi surface and the average energy of a physical system
at zero pressure (at the saturation density) are equal. This theorem is
useful for checking the internal consistency of approximations. The
mean-field theory and the relativistic Hartree approximation of the standard
Walecka model obey this theorem.

The vacuum dibaryon mass enters our model as a free parameter. The formal
assumption that the dibaryon condensate occurs at the saturation density
does not lead to any contradictions. In chemical equilibrium with respect to
the reaction $NN\leftrightarrow D$, the relations (4.16) are fulfilled. In
agreement with the HV theorem, for $p=0$ Eqs.(4.6) and (4.26) give 
\begin{equation}
\label{IV.27}\mu _N=E_F^{*}+g_\omega \omega _c=\frac \varepsilon {\rho
_{TV}}. 
\end{equation}

In Figs. 8 and 9 we show the energy per nucleon and the pressure versus the
total baryon number density for some possible dibaryons. It is useful to
compare Figs.2 and 9. In Fig.2, we schematically show the behavior of the
pressure as a function of the density in the ideal gas approximation. In
Fig.9, the pressure is calculated for the dibaryon extension of the Walecka
model. The effect of vanishing incompressibility in the ideal gas model (the
horizontal dotted line in Fig.2) is displayed in Fig.9 as a softening of the
EOS of the heterophase nuclear (dashed line in Fig.9) and neutron-dibaryon
matter (dotted line in Fig.9). The occurrence of $H$-dibaryons provides a
possible mechanism for the formation of strange matter.

Note that the pressure of the heterophase system of nucleons and dibaryons
obeys the basic inequality of statistical mechanics \cite{Kad}$\ $ 
\begin{equation}
\label{IV.28}\frac{\partial p}{\partial \rho _{TV}}\geq 0. 
\end{equation}

\subsection{The concept of equilibrium of heterophase substances for ${\cal L%
}_c=0\;$and ${\cal L}_c\neq 0$.}

We discuss here the effect of a small term ${\cal L}_c$ describing
transitions between dibaryons and nucleons.

In homophase substances, the energy density is a function of the total
baryon number density $\rho _{TV}$ only,

$$
\varepsilon =\varepsilon (\rho _{TV}). 
$$
In a heterophase substance, the energy is a function of additional
parameters $\xi _i$

$$
\varepsilon =\varepsilon (\rho _{TV},\xi _i). 
$$
The parameters $\xi _i$ characterize components of the heterophase
substance. The equilibrium state is determined by the conditions

\begin{equation}
\label{equi}\frac{\partial \varepsilon (\rho _{TV},\xi _i)}{\partial \xi _k}%
=0\;\;\;\;\;\;\;\;\;(k=1,2,...). 
\end{equation}
Eigenvalues of the matrix 
\begin{equation}
\label{posi}\frac{\partial ^2\varepsilon (\rho _{TV},\xi _i)}{\partial \xi
_k\partial \xi _l} 
\end{equation}
should be positive to guarantee a minimum of the energy.

In our case, the state is specified by the nucleon chemical potential $\mu
_N $, the dibaryon chemical potential $\mu _D$, and by the expectation value
of the dibaryon field $\rho _{DS}^c$. The nucleon chemical potential $\mu _N$
is a free parameter. The dibaryon chemical potential $\mu _D$ determines the
evolution of the condensate part of the dibaryon field (Eq.(3.16)).

The value $\mu _D$ is fixed by the Hugenholtz-Pines condition \cite{Hug}. In
the MF approximation, $\mu _D^{*}$ $=m_D^{*}$. There remain two free
parameters: $\mu _N$ and $\rho _{DS}^c$. The total density is a sum of the
two terms $\rho _{TV}=\rho _{NV}+2\rho _{DV}^c$. This expression is also
valid for ${\cal L}_c\neq 0$ (if there are no derivatives in ${\cal L}_c$).

We can chose as independent parameters the values $\rho _{NV}$ and $\rho
_{DV}^c$. It is not necessary to require that $\rho _{NV}$ and $\rho _{DV}^c$
be timelike components of two conserved currents. It is sufficient that they
characterize uniquely the phases of the binary mixture. Equations (\ref{equi}%
) give then 
\begin{equation}
\label{eqsp}2\frac{\partial \varepsilon (\rho _{NV},\rho _{DV}^c)}{\partial
\rho _{NV}}=\frac{\partial \varepsilon (\rho _{NV},\rho _{DV}^c)}{\partial
\rho _{DV}^c}. 
\end{equation}

One can use expression (\ref{IV.22}) for the energy density to verify that
Eq.(\ref{cheq}) follows from Eq.(\ref{eqsp}). This is a consequence of Eqs.(%
\ref{IV.24}) and (\ref{IV.25}). The matrix (\ref{posi}) is positively
definite when Eq.(\ref{poen}) is fulfilled.

For every conserved current one can introduce an independent chemical
potential. For ${\cal L}_c=0$, the parameters $\mu _N$ and $\mu _D$ have the
meaning of the chemical potentials corresponding to two conserved nucleon
and dibaryon currents. For ${\cal L}_c\neq 0$, there is only one conserved
baryon current and only one baryon chemical potential.

In the both cases (${\cal L}_c=0$ and ${\cal L}_c\neq 0$), the equilibrium
is determined by Eq.(\ref{eqsp}).

For ${\cal L}_c=0$, this equation reduces to Eq.(\ref{cheq})$.$ It can be
naturally interpreted in terms of the chemical equilibrium between nucleon
and dibaryon phases.

For ${\cal L}_c \neq 0$, the values $\rho _{NV}$ and $\rho _{DV}^c$ are no
longer timelike components of conserved currents. They play the role of
formal parameters $\xi _i$ characterizing the nucleon and dibaryon phases.
In such a case, the condition (\ref{eqsp}) cannot be interpreted in terms of
a chemical equilibrium, since we cannot determine individual contributions
of nucleons and dibaryons to the total baryon number of the system.

For ${\cal L}_c\neq 0$, one should write additional terms $O({\cal L}_c)$ on
the right hand side of Eqs.(4.16) and (4.17). For narrow dibaryons which we
discuss here, the corrections to these equations are small. For example, in
case of the $H$-particle the corrections are of order $10^{-10}$.

\section{GREEN'S FUNCTIONS AND ELEMENTARY EXCITATIONS}

\setcounter{equation}{0}

In the MF approximation, all dibaryons are in the Bose condensate. It is
known that interaction between Bose particles brings some fraction of bosons
out of the condensate. In the nonrelativistic approximation, the density of
bosons which are not in the condensate increases with the total density
faster than the density of bosons in the condensate \cite{Abr}.

The bosons that are not in the condensate have finite momenta and they
contribute to the pressure. One expects that the dibaryons which are not in
the condensate shift the critical density for disappearance of the effective
nucleon mass to higher values, because the dibaryon scalar charge is
suppressed by the Lorentz factor. The contribution of dibaryons out of the
condensate to the energy density and pressure can be calculated using the
diagram technique developed for Bose systems by Belyaev \cite{Bel}. To go
beyond the MF approximation, it is necessary to determine the Green's
functions of the system.

\subsection{Solutions of the Gorkov-Dyson equations}

To eliminate the time dependence from the condensate parts of the $\varphi $%
-fields, we pass to the $\mu $-representation for dibaryons. This can be
done by the substitution $\varphi \rightarrow \varphi e^{i\mu _Dt}$ and $%
\varphi ^{*}\rightarrow \varphi ^{*}e^{-i\mu _Dt}$.

The Green's functions are defined by 
\begin{equation}
\label{V.1}iG_{\alpha \beta }(x^{\prime }-x)=<T\Psi _\alpha (x^{\prime
})\bar \Psi _\beta (x)>, 
\end{equation}
\begin{equation}
\label{V.2}iD^{AB}(x^{\prime }-x)=iD^{BA}(x-x^{\prime })=<T\hat A(x^{\prime
})\hat B(x)>, 
\end{equation}
with $A$, $B=$ $\sigma $, $\omega _\mu $, $\varphi $, $\varphi ^{*}.$ In
momentum space 
\begin{equation}
\label{V.3}D^{AB}(k)=D^{BA}(-k). 
\end{equation}

The $\sigma $- and $\omega $-vertices with dibaryons are depicted in Fig.10.
The crosses on the dibaryon double lines denote the appearance or
disappearance of dibaryons from the Bose condensate. In Fig.11 we
graphically show a representation of Eqs.(4.1) and (4.2). The set of these
diagrams can be summed up either to modify the meson propagators (the bold
dashed lines) or the meson vertices with dibaryons. The dressed
meson-dibaryon vertices are determined by the diagrams shown in Fig.12.

It is useful to distinguish three kinds of propagators. The bare ones
denoted by thin lines, the mean field ones denoted by thick lines, and the
complete ones denoted by blobs with outgoing thick lines. The MF propagators
are determined by the effective nucleon and dibaryon masses (3.7) and (3.8)
and by the effective masses of the mesons (4.3) and (4.4). The graphical
representations for the MF propagators are shown in Fig.13.

The total Green's functions can be determined self-consistently by solving a
system of Gorkov-Dyson equations. As an example we derive an equation for
the $\sigma $-meson propagator. Let us multiply Eq.(3.3) by $\hat \sigma $,
take the time-ordered product of the equation, and find the average value of
the equation over the ground state

\begin{equation}
\label{V.4}
\begin{array}{c}
(-\Box -m_\sigma ^2)<T\sigma (1)\hat \sigma (2)>=\delta ^4(1,2)+g_\sigma
<T\bar \Psi (1)\Psi (1)\hat \sigma (2)> \\ 
+2h_\sigma <T(m_D+h_\sigma \sigma (1))\varphi ^{*}(1)\varphi (1)\hat \sigma
(2)>. 
\end{array}
\end{equation}
Taking into account the second-order terms with respect to the operator
fields, we obtain 
\begin{equation}
\label{V.5}
\begin{array}{c}
(-\Box -m_\sigma ^2)<T\hat \sigma (1)\hat \sigma (2)>=\delta
^4(1,2)+2h_\sigma ^2\rho _{DS}^c<T\hat \sigma (1)\hat \sigma (2)> \\ 
+(2m_D^{*}h_\sigma \sqrt{\rho _{DS}^c})(<T\hat \varphi (1)\hat \sigma
(2)>+<T\hat \varphi ^{*}(1)\hat \sigma (2)>). 
\end{array}
\end{equation}
The second term on the right hand side redefines the mass of the $\sigma $%
-meson (see Fig.13 and Eq.(4.3)). In the momentum representation, Eq.(5.5)
takes the form 
\begin{equation}
\label{V.6}D^{\sigma \sigma }(k)=\tilde D^{\sigma \sigma }(k)+\tilde
D^{\sigma \sigma }(k)2m_D^{*}h_\sigma \sqrt{\rho _{DS}^c}(D^{\varphi \sigma
}(k)+D^{\varphi ^{*}\sigma }(k)), 
\end{equation}
where 
\begin{equation}
\label{V.7}\tilde D^{\sigma \sigma }(k)=\frac 1{k^2-\tilde m_\sigma ^2}. 
\end{equation}
is the $\sigma $-meson MF propagator.

The equations for other Green's functions can be obtained in similar way to
give 
\begin{equation}
\label{V.8}D_\mu ^{\sigma \omega }(k)=\tilde D^{\sigma \sigma
}(k)2m_D^{*}h_\sigma \sqrt{\rho _{DS}^c}(D_\mu ^{\varphi \omega }(k)+D_\mu
^{\varphi ^{*}\omega }(k)), 
\end{equation}
\begin{equation}
\label{V.9}D^{\sigma \varphi }(k)=\tilde D^{\sigma \sigma
}(k)2m_D^{*}h_\sigma \sqrt{\rho _{DS}^c}(D^{\varphi ^{*}\varphi
}(k)+D^{\varphi \varphi }(k)), 
\end{equation}
\begin{equation}
\label{V.10}D^{\sigma \varphi ^{*}}(k)=\tilde D^{\sigma \sigma
}(k)2m_D^{*}h_\sigma \sqrt{\rho _{DS}^c}(D^{\varphi ^{*}\varphi
^{*}}(k)+D^{\varphi \varphi ^{*}}(k)), 
\end{equation}
\begin{equation}
\label{V.11}
\begin{array}{c}
D_{\mu \nu }^{\omega \omega }(k)=\tilde D_{\mu \nu }^{\omega \omega
}(k)+\tilde D_{\mu \tau }^{\omega \omega }(k)h_\omega \sqrt{\rho _{DS}^c}[%
(2\mu _D^{*}+k)_\tau D_\nu ^{\varphi \omega }(k) \\ +(2\mu _D^{*}-k)_\tau
D_\nu ^{\varphi ^{*}\omega }(k)], 
\end{array}
\end{equation}
\begin{equation}
\label{V.12}
\begin{array}{c}
D_\mu ^{\omega \varphi }(k)=\tilde D_{\mu \tau }^{\omega \omega }(k)h_\omega 
\sqrt{\rho _{DS}^c}[(2\mu _D^{*}+k)_\tau D^{\varphi \varphi }(k) \\ +(2\mu
_D^{*}-k)_\tau D^{\varphi ^{*}\varphi }(k)], 
\end{array}
\end{equation}
\begin{equation}
\label{V.13}
\begin{array}{c}
D_\mu ^{\omega \varphi ^{*}}(k)=\tilde D_{\mu \tau }^{\omega \omega
}(k)h_\omega \sqrt{\rho _{DS}^c}[(2\mu _D^{*}+k)_\tau D^{\varphi \varphi
^{*}}(k) \\ +(2\mu _D^{*}-k)_\tau D^{\varphi ^{*}\varphi ^{*}}(k)], 
\end{array}
\end{equation}
\begin{equation}
\label{V.14}
\begin{array}{c}
D^{\varphi \varphi ^{*}}(k)=\tilde D^{\varphi \varphi ^{*}}(k)+\tilde
D^{\varphi \varphi ^{*}}(k)[h_\omega 
\sqrt{\rho _{DS}^c}(2\mu _D^{*}+k)_\tau D_\tau ^{\omega \varphi ^{*}}(k) \\ 
+2m_D^{*}h_\sigma \sqrt{\rho _{DS}^c}D^{\sigma \varphi ^{*}}(k)], 
\end{array}
\end{equation}
\begin{equation}
\label{V.15}
\begin{array}{c}
D^{\varphi \varphi ^{*}}(k)=\tilde D^{\varphi \varphi ^{*}}(k)[h_\omega 
\sqrt{\rho _{DS}^c}(2\mu _D^{*}+k)_\tau D_\tau ^{\omega \varphi ^{*}}(k) \\ 
+2m_D^{*}h_\sigma \sqrt{\rho _{DS}^c}D^{\sigma \varphi ^{}}(k)], 
\end{array}
\end{equation}
\begin{equation}
\label{V.16}
\begin{array}{c}
D^{\varphi ^{*}\varphi ^{*}}(k)=\tilde D^{\varphi ^{*}\varphi }(k)[h_\omega 
\sqrt{\rho _{DS}^c}(2\mu _D^{*}-k)_\tau D_\tau ^{\omega \varphi ^{*}}(k) \\ 
+2m_D^{*}h_\sigma \sqrt{\rho _{DS}^c}D^{\sigma \varphi ^{*}}(k)]. 
\end{array}
\end{equation}
Here 
\begin{equation}
\label{V.17}\tilde D_{\mu \nu }^{\omega \omega }(k)=\frac{-g_{\mu \nu
}+k_\mu k_\nu /\tilde m_\omega ^2}{k^2-\tilde m_\omega ^2}, 
\end{equation}
\begin{equation}
\label{V.18}\tilde D^{\varphi \varphi ^{*}}(k)=\frac 1{(k+\mu
_D^{*})^2-m_D^{*2}} 
\end{equation}
are the $\omega $-meson and dibaryon MF propagators.

The system of equations (5.6) and (5.8)-(5.16) is shown graphically in
Fig.14. It admits an explicit solution. The propagators $D^{\varphi \varphi
^{*}}(k)$ and $D^{\varphi ^{*}\varphi ^{*}}(k)$ are expressible in terms of
the propagators $D^{\sigma \varphi ^{*}}(k)$ and $D^{\sigma \varphi ^{*}}(k)$
and analogous propagators for the $\omega $-meson. These propagators in turn
are expressible through the propagators $D^{\varphi \varphi ^{*}}(k)$ and $%
D^{\varphi ^{*}\varphi ^{*}}(k)$.

The system of two equations for dibaryon Green's functions 
\begin{equation}
\label{V.19}
\begin{array}{c}
D^{\varphi \varphi ^{*}}(k)=\tilde D^{\varphi \varphi ^{*}}(k)+\tilde
D^{\varphi \varphi ^{*}}(k)\Sigma ^{\varphi \varphi ^{*}}(k)D^{\varphi
\varphi ^{*}}(k)+\tilde D^{\varphi \varphi ^{*}}(k)\Sigma ^{\varphi \varphi
}(k)D^{\varphi ^{*}\varphi ^{*}}(k), \\ 
D^{\varphi ^{*}\varphi ^{*}}(k)=\tilde D^{\varphi ^{*}\varphi }(k)\Sigma
^{\varphi ^{*}\varphi }(k)D^{\varphi ^{*}\varphi ^{*}}(k)+\tilde D^{\varphi
^{*}\varphi }(k)\Sigma ^{\varphi ^{*}\varphi ^{*}}(k)D^{\varphi \varphi
^{*}}(k), 
\end{array}
\end{equation}
where 
\begin{equation}
\label{V.20}
\begin{array}{c}
\Sigma _{}^{\varphi \varphi ^{*}}(k)=\Sigma _{}^{\varphi ^{*}\varphi
}(-k)=(h_\omega 
\sqrt{\rho _{DS}^c})^2(2\mu _D^{*}+k)_\mu \tilde D_{\mu \nu }^{\omega \omega
}(k)(2\mu _D^{*}+k)_\nu \\ +(2m_D^{*}h_\sigma 
\sqrt{\rho _{DS}^c})^2\tilde D^{\sigma \sigma }(k), \\ \Sigma _{}^{\varphi
\varphi }(k)=\Sigma _{}^{\varphi ^{*}\varphi ^{*}}(-k)=(h_\omega 
\sqrt{\rho _{DS}^c})^2(2\mu _D^{*}+k)_\mu \tilde D_{\mu \nu }^{\omega \omega
}(k)(2\mu _D^{*}-k)_\nu \\ +(2m_D^{*}h_\sigma \sqrt{\rho _{DS}^c})^2\tilde
D^{\sigma \sigma }(k) 
\end{array}
\end{equation}
constitutes therefore a closed system of equations. These equations are
shown graphically in Fig.15. They are identical to the Gorkov equations in
the theory of superconductivity \cite{Abr}. The relativistic version of
these equations for the SU(2) color quark matter is discussed in Ref. \cite
{Gia}. The system has solutions 
\begin{equation}
\label{V.21}D^{\varphi \varphi ^{*}}(k)=\frac{\tilde D^{\varphi ^{*}\varphi
}(k)^{-1}-\Sigma ^{\varphi ^{*}\varphi }(k)}{(\tilde D^{\varphi \varphi
^{*}}(k)^{-1}-\Sigma ^{\varphi \varphi ^{*}}(k))(\tilde D^{\varphi
^{*}\varphi }(k)^{-1}-\Sigma ^{\varphi ^{*}\varphi }(k))-\Sigma ^{\varphi
^{*}\varphi ^{*}}(k)\Sigma ^{\varphi \varphi }(k)}, 
\end{equation}
\begin{equation}
\label{V.22}D^{\varphi ^{*}\varphi ^{*}}(k)=\frac{\Sigma ^{\varphi
^{*}\varphi ^{*}}(k)}{(\tilde D^{\varphi \varphi ^{*}}(k)^{-1}-\Sigma
^{\varphi \varphi ^{*}}(k))(\tilde D^{\varphi ^{*}\varphi }(k)^{-1}-\Sigma
^{\varphi ^{*}\varphi }(k))-\Sigma ^{\varphi ^{*}\varphi ^{*}}(k)\Sigma
^{\varphi \varphi }(k)}. 
\end{equation}
>From expression

\begin{equation}
\label{V.23}D^{\varphi \varphi }(k)=\tilde D^{\varphi \varphi ^{*}}(k)\Sigma
^{\varphi \varphi ^{*}}(k)D^{\varphi \varphi ^{*}}(k)+\tilde D^{\varphi
\varphi ^{*}}(k)\Sigma ^{\varphi \varphi }(k)D^{\varphi ^{*}\varphi ^{*}}(k) 
\end{equation}
we find also 
\begin{equation}
\label{V.24}D^{\varphi \varphi }(k)=\frac{\Sigma ^{\varphi \varphi }(k)}{%
(\tilde D^{\varphi \varphi ^{*}}(k)^{-1}-\Sigma ^{\varphi \varphi
^{*}}(k))(\tilde D^{\varphi ^{*}\varphi }(k)^{-1}-\Sigma ^{\varphi
^{*}\varphi }(k))-\Sigma ^{\varphi ^{*}\varphi ^{*}}(k)\Sigma ^{\varphi
\varphi }(k)}. 
\end{equation}
The Green's functions for other particles are expressible in terms of the
constructed dibaryon Green's functions.

\subsection{Dispersion laws for elementary excitations}

The mass operators $\Sigma ^{\varphi \varphi ^{*}}(k)$ has the form%
$$
\Sigma ^{\varphi \varphi ^{*}}(k)=(h_\omega \sqrt{\rho _{DS}^c})^2[\frac{%
4\mu _D^{*2}}{\tilde m_{_\omega }^2}\frac{\omega ^2-\tilde m_\omega ^2}{%
\omega ^2-{\bf k}^2-\tilde m_{_\omega }^2}+\frac{4\mu _D^{*}\omega }{\tilde
m_{_\omega }^2}+\frac{\omega ^2-{\bf k}^2}{\tilde m_{_\omega }^2}] 
$$
\begin{equation}
\label{V.25}+(h_\sigma \sqrt{\rho _{DS}^c})^2\frac{4m_D^{*2}}{\omega ^2-{\bf %
k}^2-\tilde m_\sigma ^2}. 
\end{equation}
Denoting terms appearing in this expression, successively, by $a$, $b$, $c$,
and $d$, one can write the following representations for the mass operators: 
$\Sigma ^{\varphi \varphi ^{*}}(k)=a+b+c+d$, $\Sigma ^{\varphi ^{*}\varphi
}(k)=a-b+c+d$, and $\Sigma ^{\varphi \varphi }(k)=\Sigma ^{\varphi
^{*}\varphi ^{*}}(k)=a-c+d$.

The chemical potential of the system is determined from the relation 
\begin{equation}
\label{V.26}\mu _D^{*2}-m_D^{*2}=\Sigma ^{\varphi \varphi ^{*}}(0)-\Sigma
^{\varphi \varphi }(0) 
\end{equation}
which constitutes a relativistic extension of the non-relativistic relation
derived by Hugenholtz and Pines \cite{Hug} for Bose systems. For $k=0$, $%
b=c=0$, and the right hand side of Eq.(5.26) vanishes. Therefore, the MF
relation (4.5) remains valid. A relation of such a kind is necessary to get
a pole in the dibaryon Green's functions at $\omega ={\bf k}=0$ i.e. for the
existence of sound in the medium.

The spectrum of elementary excitations of the system is determined by zeros
of the inverse Green's functions: 
\begin{equation}
\label{V.27}(\tilde D^{\varphi \varphi ^{*}}(k)^{-1}-\Sigma ^{\varphi
\varphi ^{*}}(k))(\tilde D^{\varphi ^{*}\varphi }(k)^{-1}-\Sigma ^{\varphi
^{*}\varphi }(k))-\Sigma ^{\varphi ^{*}\varphi ^{*}}(k)\Sigma ^{\varphi
\varphi }(k)=0. 
\end{equation}
To eliminate poles coming from the meson propagators, we multiply the
denominator of the Green's functions by $(k^2-\tilde m_\omega ^2)(k^2-\tilde
m_\sigma ^2)\frac{\tilde m_\omega ^2}{m_\omega ^2}$. We then get an
equivalent 4$th$ order polynomial with respect to $\omega ^2:$ 
\begin{equation}
\label{V.28}
\begin{array}{c}
\sum_{n=0}^4\omega ^{2n}c_n=(k^2-\tilde m_\omega ^2)(k^2-\tilde m_\sigma ^2) 
\frac{\tilde m_\omega ^2}{m_\omega ^2}\times \\ \lbrack (\tilde D^{\varphi
\varphi ^{*}}(k)^{-1}-\Sigma ^{\varphi \varphi ^{*}}(k))(\tilde D^{\varphi
^{*}\varphi }(k)^{-1}-\Sigma ^{\varphi ^{*}\varphi }(k))-\Sigma ^{\varphi
^{*}\varphi ^{*}}(k)\Sigma ^{\varphi \varphi }(k)] 
\end{array}
\end{equation}
with coefficients 
\begin{equation}
\label{V.29}
\begin{array}{c}
\begin{array}{c}
\begin{array}{c}
c_0= 
{\bf k}^8+{\bf k}^6(\tilde m_\omega ^2+\tilde m_\sigma ^2) \\ + 
{\bf k}^4(\tilde m_\omega ^2\tilde m_\sigma ^2+8\mu _D^{*2}h_\omega ^2\rho
_{DS}^c-8\mu _D^{*2}h_\sigma ^2\rho _{DS}^c) \\ +8 
{\bf k}^2(\tilde m_\sigma ^2\mu _D^{*2}h_\omega ^2\rho _{DS}^c-\tilde
m_\omega ^2\mu _D^{*2}h_\sigma ^2\rho _{DS}^c), \\ 
\begin{array}{c}
c_1=-4 
{\bf k}^6-{\bf k}^4(3\tilde m_\omega ^2+3\tilde m_\sigma ^2+4\mu _D^{*2}) \\ 
-2 
{\bf k}^2(\tilde m_\omega ^2\tilde m_\sigma ^2+2\tilde m_\omega ^2\mu
_D^{*2}+2\tilde m_\sigma ^2\mu _D^{*2}+4\mu _D^{*2}h_\omega ^2\rho
_{DS}^c-8\mu _D^{*2}h_\sigma ^2\rho _{DS}^c) \\ -4\tilde m_\omega ^2\tilde
m_\sigma ^2\mu _D^{*2}+8\tilde m_\omega ^2\mu _D^{*2}h_\sigma ^2\rho
_{DS}^c, 
\end{array}
\end{array}
\\ 
c_2=6 
{\bf k}^4+{\bf k}^2(3\tilde m_\omega ^2+3\tilde m_\sigma ^2+8\mu _D^{*2}) \\ 
+\tilde m_\omega ^2\tilde m_\sigma ^2+4\tilde m_\omega ^2\mu _D^{*2}+4\tilde
m_\sigma ^2\mu _D^{*2}-8\mu _D^{*2}h_\sigma ^2\rho _{DS}^c, \\ 
c_3=-4 
{\bf k}^2-\tilde m_\omega ^2-\tilde m_\sigma ^2-4\mu _D^{*2}, \\ c_4=1. 
\end{array}
\end{array}
\end{equation}

The polynomial (5.28) determines four different excitations, two of dibaryon
type (particles and antiparticles, the first excitation is sound) and two of 
$\sigma $- and $\omega $-meson types. These four poles occur in all Green's
functions because of the $\sigma $-$\omega $-$\varphi $-$\varphi ^{*}$
mixing. Such a mixing occurs because the $\sigma $- and $\omega $-mesons can
be absorbed by the dibaryons in the condensate as a result of which the
dibaryons leave the condensate and propagate as normal particles. The mixing
describes also processes with creation of dibaryon-antidibaryon pairs with
subsequent absorption of dibaryons by condensate and propagation of
antidibaryons.

One can verify that the Green's function $D^{\sigma \sigma }(k)$ has no pole
at $k^2=\tilde m_\sigma ^2.$ The poles of $D^{\sigma \sigma }(k)$ coincide
with the poles of the dibaryon Green's functions.

The second term in the $\omega $-meson propagator in Eqs.(5.11) is
factorizable and there is no $g_{\mu \nu }$ tensor structure. Therefore, the
pole at $k^2=\tilde m_\omega ^2$ in the first term cannot be cancelled by
the second term, as in case of the $\sigma $-meson. The $\omega $-meson in
heterophase nucleon-dibaryon matter has therefore two branches of
excitations.

The velocity of sound $a_s$ can be found from Eq.(5.27) by keeping terms of
order $O({\bf k}^2)$ and $O(\omega ^2)$. In this limit only the sound mode $%
\omega=\omega _s({\bf k})$ survives. We get 
\begin{equation}
\label{V.31}a_s^2=(\frac{\partial \omega _s({\bf k})}{\partial {\bf k}}%
)\left| _{{\bf k}=0}^2\right. =\frac \alpha {1+\alpha } 
\end{equation}
with 
\begin{equation}
\label{V.32}\alpha =2\rho _{DS}^c\frac{m_\sigma ^2}{\tilde m_\sigma ^2}(%
\frac{h_\omega ^2}{m_\omega ^2}-\frac{h_\sigma ^2}{m_\sigma ^2}). 
\end{equation}
When the condition for stability (3.21) is fulfilled, the value $a_s^2$ is
positive and less than unity (i.e. less than the velocity of light).

The numerical analysis of Eq.(5.27) shows that there are no complex or
negative $\omega ^2$ when the inequality (3.21) is satisfied. This result
implies the stability of the ground state of the system for $a_s^2>0$. If
the poles occur symmetrically at $\pm \;\omega _\alpha ({\bf k})$, the
denominator of the dibaryon Green's function has no zeros on the imaginary
axis of the complex $\omega $-plane. One can check that after a Wick
rotation, the denominator really becomes positive for all values of $\omega $
and ${\bf k}$.

The group velocities of all four excitations are less than the velocity of
light in absolute values. This is quite natural, because the model is
relativistically invariant. It the limit ${\bf k}\rightarrow \infty $ we can
keep the leading terms in ${\bf k}^2$ in Eq.(5.28). The polynomial (5.29)
can then be summed up to $(\omega ^2-{\bf k}^2)^4.$ The dispersion laws for
all four types of excitations behave asymptotically as $\omega ^2\sim {\bf k}%
^2$, and the group velocities approach unity in the limit of large ${\bf k}%
^2 $.

\section{DISCUSSIONS AND\ CONCLUSIONS}

The qualitative estimates based on a model of non-interacting nucleons and
dibaryons show that in normal nuclear matter a dibaryon Bose condensate does
not exist provided the inequality $m_D>1.96\;$GeV is fulfilled. A more
accurate estimate can be made on the basis of the relativistic MF model.
>From the requirement of absence of a dibaryon Bose condensate for $\rho
_{TV}\leq \rho _0$, where $\rho _0=0.15\;$ fm$^{-3}$ is the saturation
density for nuclear matter, we get for $h_\omega =2g_\omega $ a constraint 
$$
m_D>1.89\;{\rm GeV}. 
$$
This constraint is valid provided that dibaryon matter is stable against
compression. It follows that the $d_1$-resonance with a mass $m_D=1.92$ GeV
does not affect the properties of ordinary nuclei.

It would be interesting to check astrophysical data for the presence of a
dibaryon condensate in the interiors of massive neutron stars as well as
possible signatures of their instability caused by dibaryons. From the
existence of massive neutron stars, one can put a lower limit on the masses
of dibaryons. The estimates (\ref{SC}) and the results based on a more
realistic model including interactions between dibaryons \cite{Hae} show
that such a constraint can be physically significant. Conversely, the
experimental discovery of dibaryons will have important astrophysical
implications.

Phase transitions of nuclear matter to strange matter \cite{Wit,Far} have
been widely discussed in the literature (for a review see \cite{Gre}). Dense
nuclear matter with a dibaryon Bose condensate can exist as an intermediate
state below the quark-gluon phase transition. This is the case when dibaryon
matter is stable against compression. If dibaryon matter is unstable against
compression, the creation of dibaryons could be a possible mechanism for the
phase transition to quark matter.

The soft core of the $HH$-interaction \cite{Sak} is responsible for the
relatively low value of the critical density for formation of $H$-dibaryons
in nuclear and neutron matter and for the possible instability of $H$%
-matter. The energetically favorable compression of $H$-matter will
eventually lead to the formation of absolutely stable strange matter.
Possible astrophysical examples are bursters and roentgen pulsars which
accret matter from companion stars. This leads to an increase of the mass
and central density of these neutron stars. Once the density exceeds the
critical density, $H$-particles can be created, leading to the formation of
strange matter. The neutron star converts then to a strange star \cite{Oli,
BVM, Lug}.

The experimental observation of a Bose condensate of dibaryons in heavy-ion
collisions would be of great importance for understanding physics of nuclear
matter at supranuclear densities. In the center-of-mass frame of the
condensate a large fraction of dibaryons has zero velocities. When the
density decreases, dibaryons in the condensate decay to their specific
channels. Experimentalists would observe in every collision events with the
same invariant mass $m_D$ and the same total momentum. An excess of such
events can be considered as a possible signature for the formation of a
dibaryon condensate in heavy-ion collisions.

The contribution of dibaryons out of the condensate to pressure and energy
density can be calculated in relativistic Hartree approximation only. We
constructed Green's functions of the system. The one-loop calculation of the
EOS for heterophase nucleon-dibaryon matter will be given elsewhere \cite
{ABK}.\vspace{1 cm}

{\bf ACKNOWLEDGMENTS}

The authors are grateful to S. Gerasimov, M. Kirchbach and M. G. Schepkin
for useful discussions and K.F\"ohl for help in preparation of the figures.
Two of us (B. V. M. and M. I. K.) are grateful to RFBR for Grant No.
94-02-03068 and Neveu-INTAS for Grant No. 93-0023. M. I. K. acknowledges
hospitality of Institute for Theoretical Physics of University of Tuebingen,
Alexander von Humboldt-Stiftung for a Forschungsstipendium, and INTAS for
Grant No. 93-0079. This work was supported also by the BMBF under the Grant
No. 06 T\"U 746 (2).

\newpage\

\begin{center}
{\bf FIGURE CAPTIONS}
\end{center}

{\bf Fig.1.} A schematic representation for the process of dibaryon Bose
condensation in neutron matter in the ideal gas approximation. When the
chemical potential of nucleons exceeds $m_D/2$, production of dibaryons
becomes energetically favorable. Dibaryons are Bose particles. They are
accumulated in the ground state of the system with zero momentum and form a
Bose condensate.

{\bf Fig.2.} Schematic behavior of the pressure as a function of the density
in the ideal gas approximation. Above the critical density for formation of
the dibaryon Bose condensate (indicated by the arrow), the neutron chemical
potential $\mu _n$ is frozen at a value $m_D/2$ and the neutron density is
fixed, but the total baryon number density is still increasing along the
dashed line due to the dibaryon formation. Dibaryons are in the ground state
and do not contribute to the pressure. Therefore the pressure for the binary
mixture remains constant (horizontal dotted line). The solid line gives the
EOS for homophase neutron matter.

{\bf Fig.3}. Diagrams contributing to the coupling constants of the $\omega $
- and $\sigma $-mesons with dibaryons in the additive model. The dibaryon $D$
couples strongly to the $NN$-channel $(a)$, the $d^{\prime }$-dibaryon
decays into the $\pi NN$-channel $(b)$. The $\omega $-meson is coupled to
nucleons $(a,b)$, the $\sigma $-meson is coupled to nucleons and the $\pi $%
-meson in the $d^{\prime }$ $(b)$.

{\bf Fig.4}. The $\sigma $- and $\pi $-meson exchange current contributions
to the $\sigma $-meson coupling constants with the dibaryon $D$ coupled
strongly to the $NN$-channel $(a)$, and with the $d^{\prime }$-dibaryon
decaying into the $\pi NN$-channel $(b)$.

{\bf Fig.5 (a,b).} The critical density for occurrence of a Bose condensate
of dibaryons in nuclear $(a)$ and neutron $(b)$ matter versus the $\sigma $%
-meson coupling constant $h_\sigma $ for $m_D=1.88$ GeV ($=2m_N$; the long
dashed curve No. $1$), $1.96$ GeV (the solid curve No. $2$), {\it etc.} with
a step $80$ MeV. The results for the $d_1(1920)$ and $d^{^{\prime }}(2060)$
dibaryons are shown (the dashed curves). Dibaryon matter is stable against
compression when the square of the sound velocity $a_s$ is positive. This is
the case for $h_\sigma /(2g_\sigma )<0.8754$. The value $\rho _0=0.15\ $fm$%
^{-3}$ is the saturation density for nuclear matter. The occurrence of $H$%
-dibaryons in nuclear and neutron matter is denoted by the crosses.

{\bf Fig.6}. Possibilities for the behavior of the difference $2\mu _N-\mu
_D $ between the two-nucleon and the dibaryon chemical potential versus the
dibaryon fraction $2\rho _{DV}^c/\rho _{TV}$. Figure $(a)$ shows unstable
equilibrium states. Figure $(b)$ shows stable equilibrium states.

{\bf Fig.7 (a,b) }The effective nucleon mass $m_N^{*}$ in GeV versus the
dibaryon fraction $2\rho _{DV}^c/\rho _{TV}\ $in heterophase matter $(a)$.
The results do not depend on the dibaryon mass. The difference $2\mu _N-\mu
_D$ between the two nucleon chemical potentials and the dibaryon chemical
potential versus the dibaryon fraction $2\rho _{DV}^c/\rho _{TV}$ $(b)$. The
results are given for total baryon densities $1,$ $2,$ $3,$ $4,$ $5,$ and $6$
times the saturation density $\rho _0.$ The normal homophase matter is
stable when $2\mu _N-\mu _D<0$ and $\rho _{DV}^c=0$.\ An intersection of a
curve with a negative slope with the horizontal line $2\mu _N-\mu _D=0$
indicates occurrence of a stable equilibrium in heterophase matter. Two such
states for nuclear and neutron matter, occurring at $\rho _{TV}=3\rho _0$
and $\rho _{TV}=2\rho _0$, are denoted by the arrows. The results are given
for $m_D=1.96$\ GeV. The dibaryon mass does not enter the self-consistency
condition (4.11) and enters linearly in the difference $2\mu _N-\mu _D$, and
the curves for other dibaryon masses can be obtained by vertical parallel
displacements.\ The results for $m_D=2.06$ GeV ($d^{\prime }$-dibaryon) can
be obtained {\it e.g. }by a $100$ MeV negative shift, {\it etc.} The solid
lines stand for nuclear ($\gamma =4$) matter, the dashed lines stand for
neutron ($\gamma =2$) matter.

{\bf Fig.8}. The energy per nucleon in homophase nuclear and neutron matter
(solid lines) and in heterophase matter (dashed and dotted lines) versus the
total baryon number density $\rho _{TV}=\rho _{NV}+2\rho _{DV}^c$for $%
d_1(1920)$ and $d^{^{\prime }}(2060)$ dibaryons using $h_\omega =2g_\omega $%
and $h_\sigma /(2g_\sigma )=0.8$. The dibaryon Bose condensation decreases
the energy of the ground states. The occurrence of $H$-particles in nuclear
and neutron matter is shown. It results in the formation of strange matter.
The value $\rho _0$ is the saturation density of nuclear matter.

{\bf Fig.9}. Equation of state for homophase nuclear and neutron matter
(solid lines) and for heterophase nuclear- and neutron-dibaryon matter
(dashed and dotted lines) versus the total baryon number density for $%
d_1(1920)$ and $d^{^{\prime }}(2060)$ dibaryons at $h_\omega =2g_\omega $and 
$h_\sigma /(2g_\sigma )=0.8$. Dibaryon Bose condensation at high densities
softens the EOS for nuclear and neutron matter. The critical density for the
occurrence of $H$-particles in nuclear and neutron matter is indicated.

{\bf Fig.10}. There are two kinds of vertices corresponding to interactions
of the $\omega $- and $\sigma $-mesons with dibaryons and a vertex
describing creation and absorption of dibaryons by the condensate.

{\bf Fig.11}. Pictorial representation of the series for the $\omega $- and $%
\sigma $-mesons mean fields (Eqs.(4.1) and (4.2) ). The diagrams can be
summed up to produce $(i)$the dressed meson MF propagators without
modification of the meson vertices or $(ii)$the dressed meson MF vertices
with dibaryons without modification of the meson MF propagators and the
meson-nucleon MF vertices.

{\bf Fig.12}. The dressed $\omega $- and $\sigma $-meson MF vertices with
dibaryons.

{\bf Fig.13}. The Dyson equations for the MF propagators of the $\sigma $-
and $\omega $-mesons, nucleons, and dibaryons. Thin lines define the bare
propagators, thick lines define the MF propagators. The dashed lines, the
solid lines, and the double solid lines describe, respectively, the meson
propagators, the nucleon propagator, and the dibaryon propagator.

{\bf Fig.14}. Pictorial representation of the Gorkov-Dyson equations for the
complete Green's functions in the heterophase nucleon-dibaryon matter. Note
the correspondence between the diagrams and the equations in the text: $(a)$%
- Eqs.(5.6) and (5.11), $(b)$- Eq.(5.8), $(c)$- Eqs.(5.9) and (5.12), $(d)$-
Eqs.(5.10) and (5.13), $(e)$- Eq.(5.14), $(f)$- Eq.(5.16), and $(g)$%
-Eq.(5.15).

{\bf Fig.15}. Pictorial representation of the system of two coupled
equations (5.19) for the normal and anomalous dibaryon Green's functions.

\end{document}